\documentclass[prb,twocolumn,showkeys,amsmath,amssymb]{revtex4-1}
\usepackage{latexsym}
\usepackage{graphicx}
\usepackage{times}
\usepackage{color}
\usepackage{amsmath}
\usepackage{dcolumn}
\usepackage{amssymb,bm,euscript}
\usepackage{bm}
\usepackage{tabularx}
\usepackage{float}

\begin{document}



\title{Alloying-induced topological transition in 2D transition-metal dichalcogenide semiconductors}

\author{Liying Ouyang, Ge Hu, Can Qi and Jun Hu}
\email[]{E-mail: jhu@suda.edu.cn}
\affiliation{School of Physical Science and Technology \& Jiangsu Key Laboratory of Thin Films, Soochow University, Suzhou 215006, China.}

\begin{abstract}
Research on two-dimensional (2D) topological insulators (TIs) is obstructed due to the lack of feasible approach to grow 2D TIs in experiment. Through systematic first-principles calculations and tight-binding simulations, we proposed that alloying Os in 2D MoX$_2$ (X$=$S, Se, Te) monolayers is an effective approach to induce semiconductor-to-TI transition, with sizable nontrivial gap of 25$\sim$37 meV. Analysis of the electronic structures reveals that the topological property mainly originates from the $5d$ orbitals of Os atom. Furthermore, the TI gaps can be modulated by external biaxial strain.
\end{abstract}



\maketitle


Topological insulator (TI) state, discovered in recent years, is a new state of condensed matters. \cite{KaneMele1, KaneMele2, ZhangBernevig1, HgTeTheory, HgTeExp, KaneReview, QiReview} It is characterized by the combination of insulating bulk state and quantized helical conducting edge state, which exhibits intriguing quantum spin Hall (QSH) effect. The helical state provides intrinsic spin lock localized at the edge and is robust against elastic backscattering, so it is ideal for various applications that require dissipationless spin transport. \cite{KaneReview, QiReview} Although QSH state was firstly predicted in graphene which is an ideal two-dimensional (2D) material, it is difficult to observe the QSH effect in graphene due to the weak intrinsic spin-orbit coupling (SOC). \cite{Hu-1, Hu-2, Hu-3} On the contrary, significant progress on the investigation of TIs in experiment was made in quantum wells \cite{HgTeTheory, HgTeExp, QW-1, QW-2} and three-dimensional (3D) TIs \cite{BiSb-theory, BiSb-exp, Bi2Se3-theory, Bi2Se3-exp1}. However, the transport properties in 3d TIs are more difficult to control than in 2D TIs, due to the gapless side surface states in 3D TIs. Therefore, 2D TIs are more promising for practical applications, since the quantized helical states only exist at the edges.

Many possible 2D TIs have been predicted in theory recently, \cite{WengHM1, Bansil-1, HeineT, ZhaoMW} but most predicted 2D TIs must be decorated by anion atoms (such as H and halogen elements) and kept away from substrate. However, it is difficult to grow freestanding 2D TIs in experiment. For example, freestanding honeycomb silicene was predicted to be TI with nontrivial gap of 1.55 meV, but the honeycomb silicene has been fabricated only on some transition metal surfaces such as Ag(111) and Ir(111). \cite{Silicene-1, Silicene-2, Silicene-3} Unfortunately, there is no evidence yet that the honeycomb silicene on these transition metal surfaces preserves the QSH state. Furthermore, large TI gaps are desired for possible applications of the 2D TIs at room temperature. Therefore, it is urgent and important to search for 2D TIs that not only have large TI gaps but also are feasible to fabricate.

Recently, 2D transition metal dichalcogenide (TMD) monolayers have attracted great attention, because they exhibit versatile electronic properties yet are chemically and physically stable. \cite{TMD-1,TMD-2} Most importantly, the 2D TMD monolayers can be obtained easily either through exfoliation of bulk materials or bottom-up syntheses. \cite{TMD-3,TMD-4} In this family, the 2D molybdenum and tungsten dichalcogenide monolayers --- MoX$_2$ and WX$_2$ (X = S, Se or Te) --- are semiconductors with sizable band gaps $\sim1$ eV and possess fascinating valleytronic character. \cite{TMD-5} Interestingly, QSH states were predicted in 2D MoX$_2$ and WX$_2$ monolayers with either structural distortion \cite{LiJu} or in metastable phase \cite{FangZhong}. These studies provide new opportunities to explore the QSH effect in semiconducting 2D materials.

Alloying with exotic elements in semiconductors has been demonstrated as an effective method to engineer its electronic property, \cite{My-JACS,ChengYC,MoWSe2,MoWS2} which may induce intriguing physical and chemical features such as dilute magnetism \cite{ChengYC} or adjustable SOC strength \cite{MoWSe2}. Therefore, it is possible to produce QSH states in 2D MoX$_2$ and WX$_2$ by alloying. Meanwhile, the QSH states are closely associated with the strength SOC, hence strong SOC is desired to achieve large TI gap. In this paper, we investigated the electronic properties of 2D MoX$_2$ and WX$_2$ monolayers alloyed with 5d transition metal elements, through first-principles calculations and tight-binding simulations. We found that Os is a good candidate to turn MoX$_2$ and WTe$_2$ into TIs, with nontrivial band gaps ranging from 5.3 meV to 32.3 meV. Moreover, the TI gaps can be tuned by external strain.


The structural and electronic properties were calculated with density functional theory (DFT) as implemented in the Vienna {\it ab-initio} simulation package. \cite{VASP1, VASP2} The interaction between valence electrons and ionic cores was described within the framework of the projector augmented wave (PAW) method. \cite{PAW1,PAW2} The generalized gradient approximation (GGA) was used for the exchange-correlation potentials and the SOC effect was invoked self-consistently. \cite{PBE} The energy cutoff for the plane wave basis expansion was set to 500 eV. A 2$\times$2 supercell was employed and the 2D Brillouin zone was sampled by a 27$\times$27 k-grid mesh. The atomic positions were fully relaxed with a criterion that requires the forces on each atom smaller than 0.01 eV/{\AA}.
The band topology is characterized by the topological invariant $\mathbb{Z}_2$, with $\mathbb{Z}_2=1$ for TIs and $\mathbb{Z}_2=0$ for ordinary insulators. \cite{Z2}  We adopted the so-called $n-$field scheme to calculate $\mathbb{Z}_2$. \cite{n-field-0, n-field-1, n-field-2}. In addition, the band structure of the one-dimensional (1D) nanoribbon of a 2D TI exhibits edge bands within its nontrivial gap. We used the Wannier90 code \cite{Wannier-3} to obtained the hopping parameters and the tight-binding model developed by Wu $et~al.$ \cite{WannierTools} to calculate the band structures of 1D nanoribbons.


\begin{figure}
\centering
\includegraphics[width=7cm]{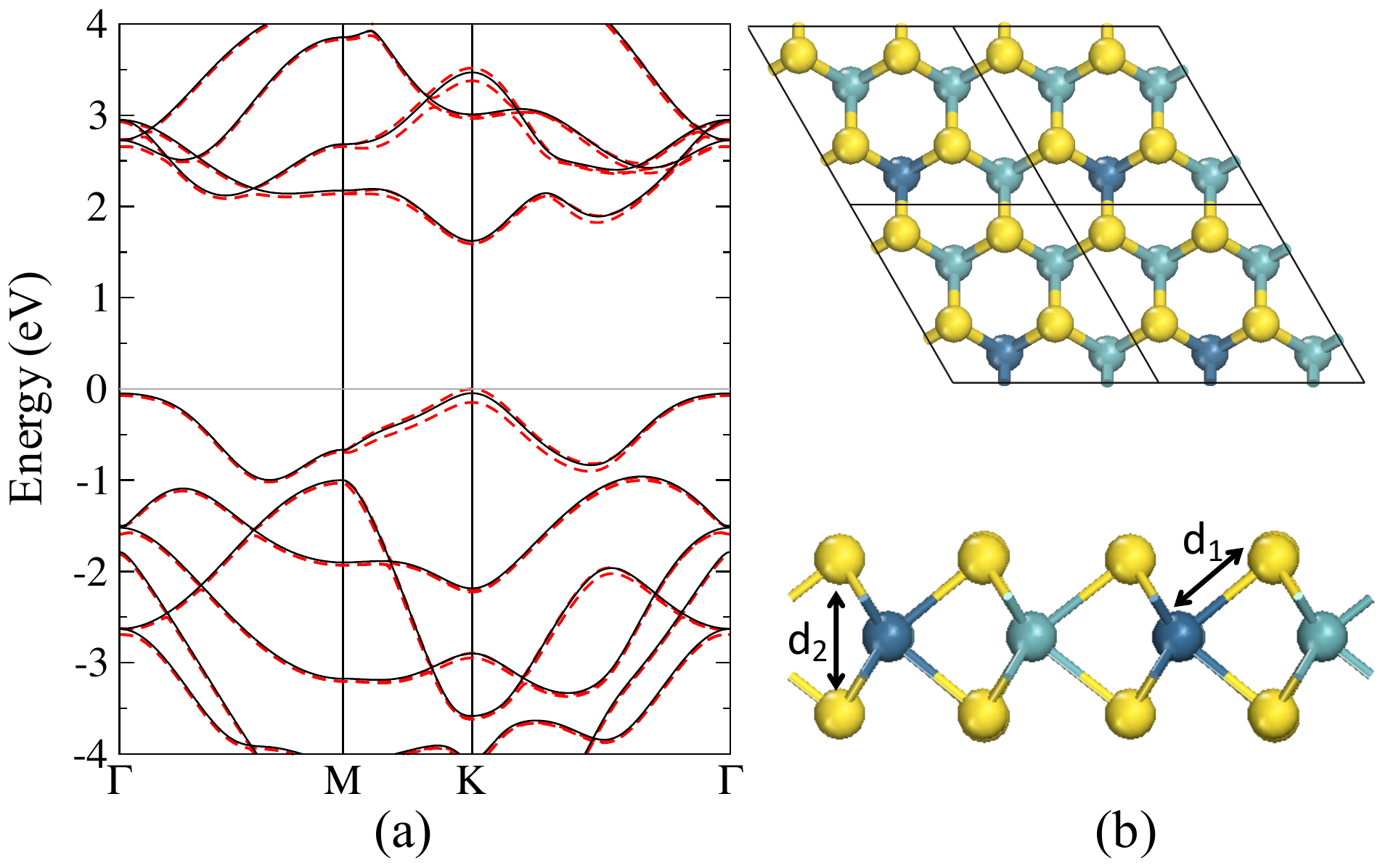}
\caption{(Color online) (a) Band structure of MoS$_2$ monolayer without (black solid curves) and with (red dashed curves) SOC. The grey horizontal line indicates the Fermi level. (b) Top and side views of the Os alloyed MX$_2$ monolayer with a $2\times2$ supercell of pure MX$_2$. The dark cyan, light cyan and yellow spheres represent Os, M and X atoms.
}
\label{structure}
\end{figure}

\begin{table}
 \centering
 \caption{The lattice constants ($a$, in {\AA}) and band gaps ($E_g$, in eV) of MX$_2$ (M = Mo, W; X = S, Se, Te) monolayers. The SOC has been involved. 
}
 \tabcolsep0.12in             
 \begin{tabular}{cccccccc}
   \hline
   \hline
   &  \multicolumn{3}{c}{Mo} & & \multicolumn{3}{c}{W} \\
    \cline{2-4} \cline{6-8}
    & S & Se & Te & & S & Se & Te \\
      \hline
   $a$ & 3.19 & 3.33 & 3.56 & & 3.19 & 3.33 & 3.56 \\
   $E_g$ & 1.59 & 1.31 & 0.94 & & 1.53 & 1.24 & 0.75 \\
   \hline
   \hline
  \end{tabular}
\end{table}

We firstly optimized the lattice constants and calculated the band structures of the Mo and W dichalcogenide monolayers. As listed in Table I, the lattice constants are slightly larger than those of their bulk counterparts \cite{TMD-1} but in agreement with previous calculations. \cite{Ding-TMD} In addition, they increase as the anion changes from S to Te owing to the increasing atomic radii of the anions. However, the lattice constants do not differ visibly for MoX$_2$ and WX$_2$ monolayers with the same anion element. On the other hand, all MoX$_2$ and WX$_2$ monolayers are semiconductors with sizable band gaps ($E_g$) as seen in Table I. The gaps undergo an indirect-to-direct transition from the bulk to monolayer, and are enlarged due to the quantum confinement in monolayer. \cite{TMD-1,TMD-3} In addition, the calculated gaps are smaller than the experimental values. For instance, the calculated $E_g$ of MoS$_2$ monolayer is 1.59 eV, while the experimental value is 1.8 eV. \cite{TMD-3} This is caused by the well known problem that DFT calculations usually underestimate band gaps of semiconductors. Nevertheless, the band structure plotted in Figure~\ref{structure}a capture the main electronic feature of the MoS$_2$ monolayer, that is, the MoS$_2$ monolayer is a direct-band-gap semiconductor with the valence band maximum (VBM) and conduction band minimum (CBM) locating at K point. Moreover, the SOC effect results in splitting of the valence bands around the K point as well as $-$K point. Owing to the time reversal symmetry, the splitting leads to fascinating valley Hall effect known as valleytronics in these monolayers. \cite{TMD-5}

\begin{figure}
\centering
\includegraphics[width = 8.5 cm]{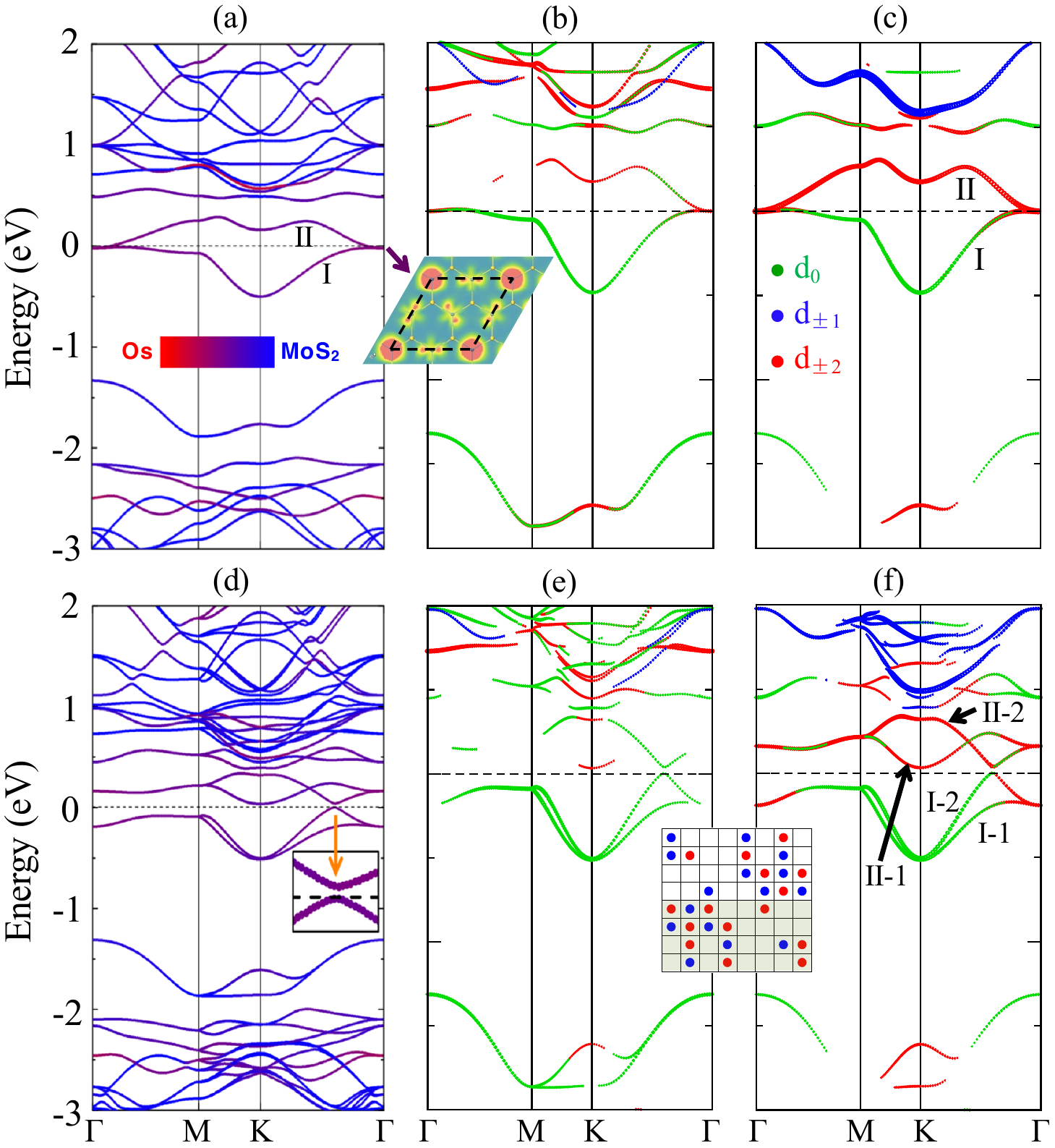}
\caption{(Color online) Band structures of Mo$_{0.75}$Os$_{0.25}$S$_2$ alloy without (a$-$c) and with (d$-$f) SOC. The horizontal dashed lines indicate the Fermi level. (a,d) Projection on Os and MoS$_2$. The color bar represents the weight of the Os atom and the host MoS$_2$.
(b,e) Projection on Mo 4d orbitals. (c,f) Projection on Os 5d orbitals. For simplicity, the $d_{z^2}$, $d_{xz/yz}$ and $d_{xy/x^2-y^2}$ orbitals are notated as $d_0$, $d_{\pm1}$ and $d_{\pm2}$, respectively. The sizes of the dots stand for the weights of the corresponding orbitals. The inset in (b) displays the charge density projected on the cation plane of the energy level at $\bf\Gamma$ point as indicated by the purple arrow.
 The inset in (e) and (f) shows the $n-$field configuration. The nonzero points are denoted by red ($n=1$) and blue ($n=-1$) dots, respectively. The $\mathbb{Z}_2$ invariant is obtained by summing the $n-$field over half Brillouin zone marked by the shadow.
}
\label{MoS2}
\end{figure}

It has been proven that doping or alloying with other transition-metal elements in MoS$_2$ is an effective way to engineer the electronic property. \cite{MoWSe2,MoWS2,TM-dope} Therefore, we choose MoS$_2$ as a prototype to investigate the effect of alloying with 5d transition-metal elements on the electronic structures. One of the Mo atoms in a 2$\times$2 supercell is replaced by a 5d transition-metal atom, resulting in an alloyed compound with concentrations of Mo and the incorporated 5d transition-metal element as 75\% and 25\%, respectively, as shown in Figure~\ref{structure}b. We considered a series of 5d transition-metal elements from Ta to Ir, and found that alloying with W does not lead to visible change of the band structures, because MoS$_2$ and WS$_2$ have almost the same atomic structure and close band structure. \cite{MoWSe2,MoWS2} Alloying with Ta, Re and Ir results in metallic property. Interestingly, MoS$_2$ alloyed with Os (notated as Mo$_{0.75}$Os$_{0.25}$S$_2$) shows a SOC induced band gap near the middle point of the path from $\bf\Gamma$ to {\bf K}, which is commonly a signature of QSH states. Accordingly, we focus on Os alloyed compounds in the following.

\begin{table}
 \centering
 \caption{Properties of M$_{0.75}$Os$_{0.25}$X$_2$ (M = Mo, W; X = S, Se, Te) monolayers: the lattice constants ($a$, in {\AA}), global and direct band gaps ($E_g$ and $E_g^{\prime}$, in eV), $\mathbb{Z}_2$ invariant, and formation energy (in eV) ($\Delta H_f = E (Alloy) -\mu_{Os} - E (MX_2) + \mu_{M}$ where $\mu$ stands for the chemical potential.). The SOC has been involved.}
 \tabcolsep0.12in             
 \begin{tabular}{cccccccc}
   \hline
   \hline
   &  \multicolumn{3}{c}{Mo} & & \multicolumn{3}{c}{W} \\
    \cline{2-4} \cline{6-8}
    & S & Se & Te & & S & Se & Te \\
      \hline
   $a$ & 6.50 & 6.78 & 7.27 & & 6.48 & 6.78 & 7.26 \\
   $E_g$ & 32.3 & 31.8 & 25.4 & & - & - & 5.3 \\
   $E_g^{\prime}$ & 37.4 & 31.8 & 25.4 & & 29.6 & 8.8 & 5.3 \\
   $\mathbb{Z}_2$ & 1 & 1 & 1 & & - & - & 1 \\
   $\Delta H_f$ & 3.53 & 2.98 & 2.02 & & 3.55 & 2.85 & 1.55 \\
   \hline
   \hline
  \end{tabular}
\end{table}

The optimized lattice constant of Mo$_{0.75}$Os$_{0.25}$S$_2$ alloy is listed in Table II, which expands by 1.9\% compared to that of MoS$_2$ due to the larger atomic size of Os than Mo. To reveal the electronic property of this alloy, we plotted the atom- and orbital-resolved band structures in Figure~\ref{MoS2}. From Figure~\ref{MoS2}a, it can be seen that the Os atom hybridizes strongly with the host MoS$_2$, which induces two bands (`I' and `II') in the gap of MoS$_2$. Without including the SOC effect, these bands are degenerate at the $\bf\Gamma$ point and retain atomic orbital character as represented by the corresponding local charge density in the inset of Figure~\ref{MoS2}. Clearly, the charge density has local $C_{3v}$ symmetry around the Os atoms, and characterizes apparent in-plane components of the $d$ orbitals (i.e. $d_{xy}$ and $d_{x^2-y^2}$) of Mo and Os atoms. Note that the charge density around the Os atom has round shape, mainly due to the large sizes of the $d_{xy}$ and $d_{x^2-y^2}$ orbitals which hybridize strongly with the S atoms. Accordingly, we sort the $d$ orbitals into three groups: (i) $d_{z^2}$; (ii) $d_{xz}$ and $d_{yz}$; and (iii) $d_{xy}$ and $d_{x^2-y^2}$. From the projections of these orbitals on the electronic bands in Figure~\ref{MoS2}b and~\ref{MoS2}c, we can see that the energy level at the $E_F$ of the $\bf\Gamma$ point originates from the $d_{xy/x^2-y^2}$ orbitals of both Mo and Os atoms and the weight of the Os atom is significantly larger than that of the Mo atom. As the bands propagate from the $\bf\Gamma$ point to the {\bf M} or {\bf K} points, the degeneracy at the $\bf\Gamma$ point breaks and the energy level evolves into two bands (notated as `I' and `II' in Figure~\ref{MoS2}c). The band `I' goes downwards, with the weight of the $d_{xy/x^2-y^2}$ orbitals decreasing and the weight of the $d_{z^2}$ orbital increasing. Finally, the band `I' near the {\bf K} point is contributed by the $d_{z^2}$ orbital completely, and the weight of the Os atom is slightly larger than that of the Mo atom. The band `II' goes upwards and keeps pure $d_{xy/x^2-y^2}$ state of the Os atom. For the $d_{xz/yz}$ orbitals, those of the Os atom contribute to the bands $\sim0.5$ eV above the $E_F$, while those of the Mo atom do not have notable contribution to the bands within the energy range in Figure~\ref{MoS2}b.

When the SOC effect is included, the degeneracy of the bands `I' and `II' at the $\bf\Gamma$ point is removed with a large splitting of 0.45 eV, as shown in Figure~\ref{MoS2}d. Each of these bands further splits into two bands along the path of $\bf\Gamma-\bf{K}-\bf{M}$. Interestingly, the band `I-2' goes upwards faster than the band `I-1' from {\bf K} to $\bf\Gamma$, so it crosses the $E_F$. Meanwhile, the band `II-2' goes downwards and crosses the $E_F$ at the same $k$ point. As a consequence, band inversion occurs between the bands `I-2' and `II-2', i.e. the $d_{z^2}$ and $d_{xy/x^2-y^2}$ orbitals, which is a signature of nontrivial topological phase in this material. \cite{HgTeTheory} As a consequence, the bands `I-2' and `II-2' interact with each other through the SOC Hamiltonian, yielding a nontrivial band gap of 37.4 meV. Note that the actual CBM (i.e. the minimum of the bands `II-1' and `II-2' as seen in Figure~\ref{MoS2}f) locates at {\bf K} point rather than the crossing point, which suggests that the global gap of Mo$_{0.75}$Os$_{0.25}$S$_2$ is actually an indirect band gap with a smaller amplitude of 32.3 meV. It is well known that GGA calculations usually underestimate band gaps of semiconductors, but hybrid functionals such as HSE \cite{HSE06} can predict more accurate band gaps. Indeed, our HSE calculations for Mo$_{0.75}$Os$_{0.25}$S$_2$ show that the nontrivial band gap can be as large as 116 meV. Furthermore, the $E_F$ locates in this SOC induced gap. Accordingly, we can conclude that the Mo$_{0.75}$Os$_{0.25}$S$_2$ alloy is a natural TI which does not require the gate voltage to adjust the $E_F$. This is a good feature for both experimental investigations and practical applications.

\begin{figure}
\centering
\includegraphics[width = 7 cm]{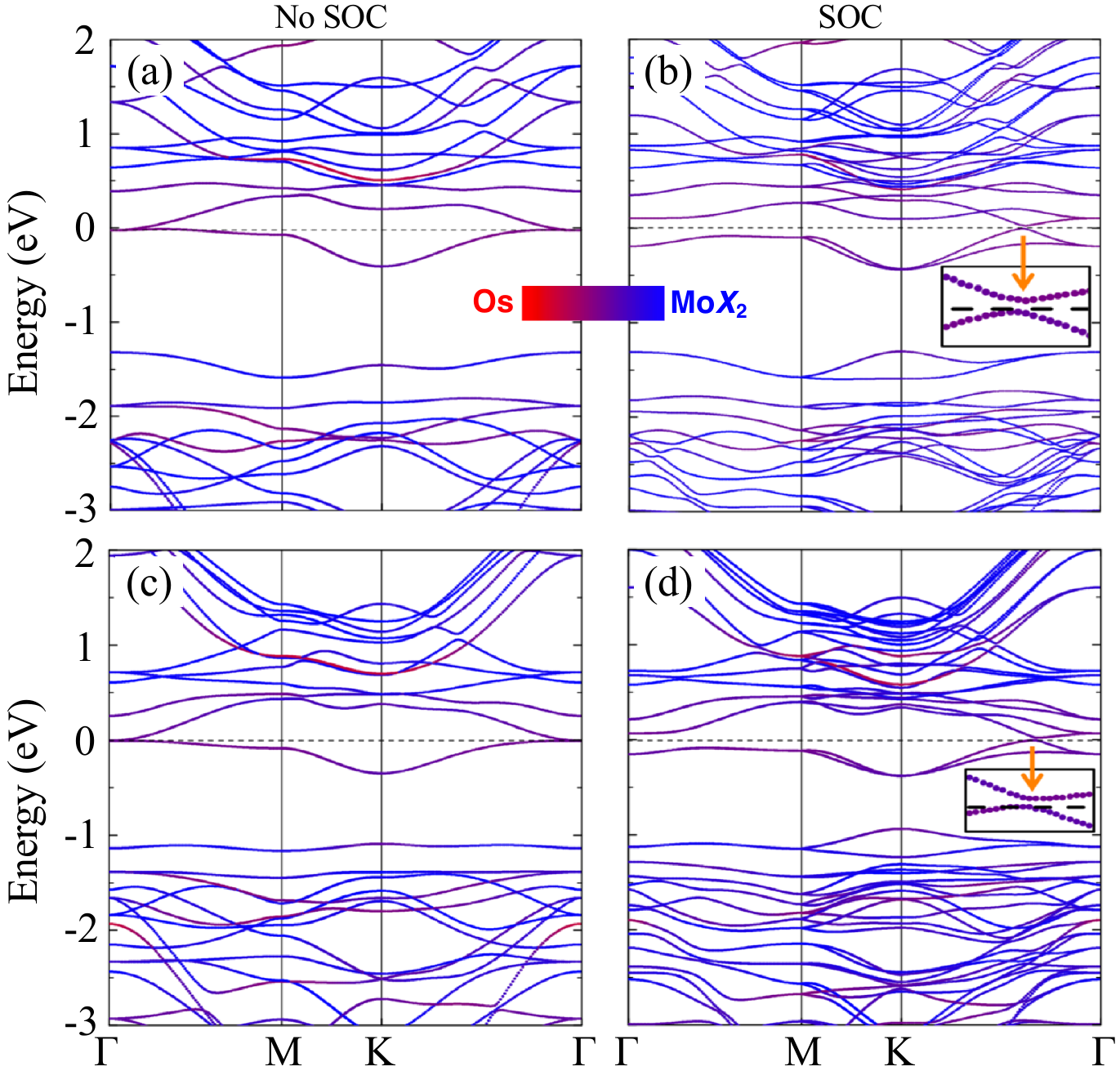}
\caption{(Color online) Species-resolved band structures of Mo$_{0.75}$Os$_{0.25}$Se$_2$ (a,b) and Mo$_{0.75}$Os$_{0.25}$Te$_2$ (c,d) without and with SOC, respectively. The horizontal dashed line represents the Fermi level.
}
\label{Se-Te}
\end{figure}


For MoSe$_2$ and MoTe$_2$, alloying with Os also leads to lattice expansion as listed in Table II similar to MoS$_2$. The band structures before and after including the SOC effect are plotted in Figure~\ref{Se-Te}. Clearly, the SOC has two effects on the electronic band structures. Firstly, it eliminates the degeneracy of the two bands beside the $E_F$ at the $\bf\Gamma$ point. Secondly, it induces band inversion when the bands propagate from $\bf\Gamma$ to {\bf K} for both Mo$_{0.75}$Os$_{0.25}$Se$_2$ and Mo$_{0.75}$Os$_{0.25}$Te$_2$, which results in nontrivial band gaps of 31.8 and 25.4 meV, respectively. Note that both gaps are direct band gaps, with the VBM and CBM at the same $k$ point in the Brillouin zone.
The band structures of W$_{0.75}$Os$_{0.25}$X$_2$ are similar to Mo$_{0.75}$Os$_{0.25}$Te$_2$. However, the `II-1' band of both W$_{0.75}$Os$_{0.25}$S$_2$ and W$_{0.75}$Os$_{0.25}$Se$_2$ decreases significantly and crosses the $E_F$, which turns the alloys into metallic. In contrast, W$_{0.75}$Os$_{0.25}$Te$_2$ still possess a nontrivial band gap of 5.3 meV with $\mathbb{Z}_2=1$. Furthermore, the direct band gap at the band-inversion point between the `I-2' and `II-2' bands still follow the same trend as Mo$_{0.75}$Os$_{0.25}$X$_2$, as listed in Table II.

\begin{figure}
\centering
\includegraphics[width = 8.5 cm]{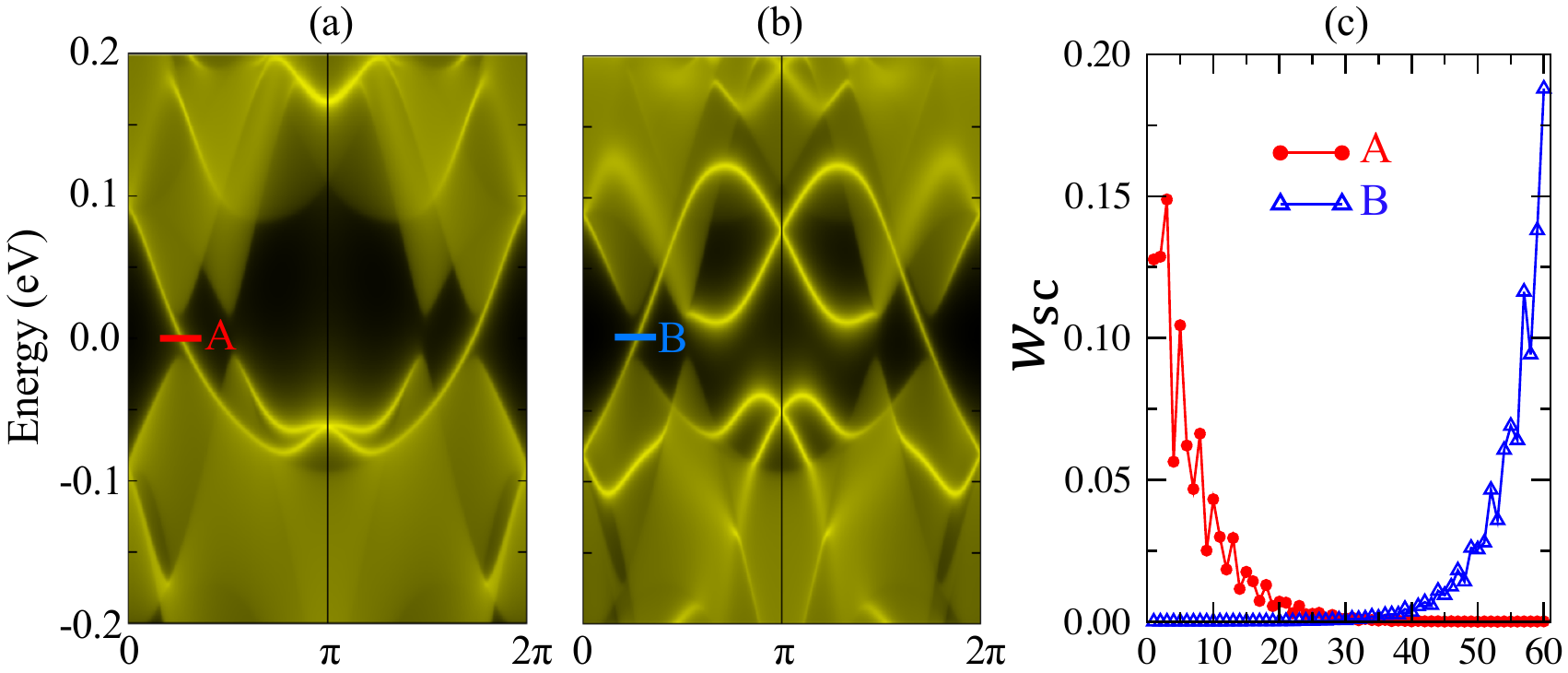}
\caption{(Color online) Edge states of the 1D Mo$_{0.75}$Os$_{0.25}$Se$_2$ nanoribbon. (a) and (b) Density of states of  on left and right edges, respectively. (c) Weight of each supercell ($w_{sc}$) from one edge to the other (labeled from 1 to 60) to the wavefunctions of the states marked as `A' and `B' in (a) and (b).
}
\label{edge}
\end{figure}

The electronic band topology can be characterized by the $\mathbb{Z}_2$ invariant. So we calculated $\mathbb{Z}_2$ of the Mo$_{0.75}$Os$_{0.25}$X$_2$ alloys with the $n-$field method. \cite{n-field-0, n-field-1, n-field-2} By counting the positive and negative $n-$field numbers over half of the torus as indicated in the inset in Figure~\ref{MoS2}, we obtained $\mathbb{Z}_2=1$ for the Mo$_{0.75}$Os$_{0.25}$S$_2$ alloy, clearly demonstrating the nontrivial band topology. Similarly, the $\mathbb{Z}_2$ invariant of the other cases is also 1, as listed in Table II. On the other hand, the topological insulators are also manifested by quantized edge states which bring about the QSH effect. \cite{KaneReview, QiReview} Therefore, we calculated the band structure of 1D nanoribbons of Mo$_{0.75}$Os$_{0.25}$X$_2$ with zigzag edges. The nanoribbons consist of 60 cells and the outermost three cells are treated as edge region. Figure~\ref{edge}a and ~\ref{edge}b display the edge density of states of 1D Mo$_{0.75}$Os$_{0.25}$Se$_2$ nanoribbon. Apparently, there are linearly dispersive bands in the gap of 2D Mo$_{0.75}$Os$_{0.25}$Se$_2$, and they are contributed by either left edge or right edge. In addition, the bands from the two edges are different, because one edge is ended by Se atoms while the other edge by Mo/Os atoms. To determine how wide the edge bands distribute, we projected the wavefunctions of the edge bands (marked as `A' and `B') on each cell and plotted the weights in Figure~\ref{edge}c. Clearly, the edge bands mainly localize within the outermost three cells and decay to zero until the thirtieth cell. Furthermore, the edge bands between $0$ and $\pi$ are from majority spin channel, while the edges bands between $\pi$ and $2\pi$ are from minority spin channel. Therefore, there are two types of electrons with opposite spins on each edge and they are propagating along opposite directions, which are the main features of the QSH effect.

\begin{figure}
\centering
\includegraphics[width=6.5cm]{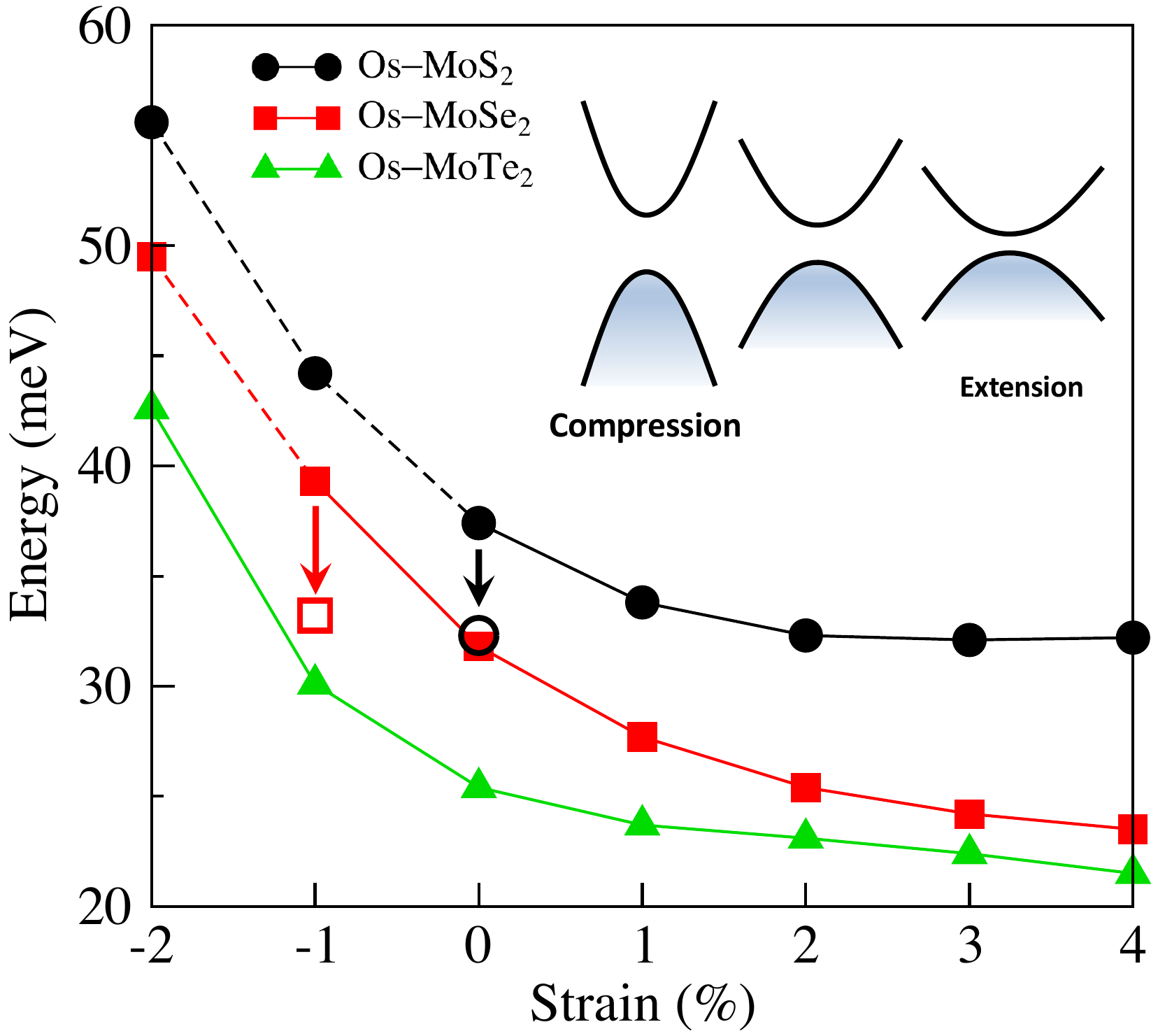}
\caption{(Color online) The nontrivial direct band gaps as a function of biaxial strain for Mo$_{0.75}$Os$_{0.25}$X$_2$ (X = S, Se and Te). 
The open square and circle indicate the indirect band gaps which are smaller the direct band gaps. The dashed lines denote that the systems become metallic. The inset shows the schematic band evolutions under compressive and extensile strains.
}
\label{strain}
\end{figure}

From table II, we can see that the direct band gaps of both Mo$_{0.75}$Os$_{0.25}$X$_2$ and W$_{0.75}$Os$_{0.25}$X$_2$ decrease as the atomic sizes of the anions increase. This is because when the Os-X bond lengths increase, the interaction between the bands `I-2' and `II-2' (Figure~\ref{MoS2}f) weakens. Therefore, the band gap may be engineered by external strain, as indicated by the inset in Figure~\ref{strain}. We calculated the band structures of Mo$_{0.75}$Os$_{0.25}$X$_2$ alloys under biaxial strain from -2\% to 4\%, and found that the bands near the direct gaps indeed undergo similar evolutions indicated in the inset in Figure~\ref{strain}. As plotted in Figure~\ref{strain}, the nontrivial direct band gaps decrease slightly under extensile strain while increase significantly under compressive strain. For Mo$_{0.75}$Os$_{0.25}$S$_2$, the compressive strain leads to metallic electronic property, because the `II-1' band decreases significantly and crosses the $E_F$. For Mo$_{0.75}$Os$_{0.25}$Se$_2$, a compressive strain of -1\% turns the gap into indirect band gap (33.2 meV), slightly larger than that without strain. When the compressive strain increases to -2\%, Mo$_{0.75}$Os$_{0.25}$Se$_2$ becomes metallic. The band gap of Mo$_{0.75}$Os$_{0.25}$Te$_2$ keeps direct under compressive strain up to -2\% and increases to 42.6 meV, 1.7 times larger than that without strain (25.4 meV). Therefore, compressive strain is an effective way to engineer the nontrivial band gap of Mo$_{0.75}$Os$_{0.25}$Te$_2$.


Beside the electronic properties, the structure stability is also an important issue for the practical fabrication of the Mo$_{0.75}$Os$_{0.25}$X$_2$ and W$_{0.75}$Os$_{0.25}$X$_2$ alloys, which can be estimated by calculating the formation energy ($\Delta H_f$). As listed in Table II, the formation energies of all the considered cases are positive, and decrease from S to Te for both MoX$_2$ and WX$_2$. Therefore, these alloys could not be produced in equilibrium growth condition. Nevertheless, the amplitudes of the formation energies are relatively small, especially for Mo$_{0.75}$Os$_{0.25}$Te$_2$ and W$_{0.75}$Os$_{0.25}$Te$_2$ (2.02 and 1.55 eV, respectively), so they may be obtained with delicate non-equilibrium growth conditions. In fact, non-equilibrium growth processes are widely used to fabricate metastable states of materials. \cite{nonequilibrium1,nonequilibrium2}


In summary, we predicted that alloying the 2D MoX$_2$ and WX$_2$ monolayers with Os leads to semiconductor-to-TI transition and produces sizable nontrivial band gaps of 5.3$\sim$32.3 meV, based on systematic first-principles calculations and tight-binding modeling. In these alloys, the band inversion occurs between the $d_{z^2}$ and $d_{xy/x^2-y^2}$ orbitals of the Os atom caused by the SOC effect, which plays the most important role in the QSH state. Interestingly, the TI gaps can be tuned by external biaxial strain. In particular, the TI gap of Mo$_{0.75}$Os$_{0.25}$Te$_2$ can be enhanced to 42.6 meV with compressive biaxial strain of -2\%. It is worth pointing that the band gaps were largely underestimated by GGA calculations, e. g. the nontrivial band gap of Mo$_{0.75}$Os$_{0.25}$S$_2$ is 116 meV from HSE calculation, greatly larger than 32.3 meV (GGA). Since 2D WX$_2$ monolayers can be fabricated in experiment effectively, our study paves the way to engineer QSH states in 2D semiconductors.

{\bf Acknowledgements} This work is supported by the National Natural Science Foundation of China (11574223), the Natural Science Foundation of Jiangsu Province (BK20150303) and the Jiangsu Specially-Appointed Professor Program of Jiangsu Province.



\end{document}